\documentclass[11pt,preprint]{JHEP}
 \preprint{26/05/2012. V2: 05/07/2012}

\title{Supermembrane interaction with dynamical D=4 N=1 supergravity.
Superfield Lagrangian description  and spacetime equations of motion.}

\author{Igor A. Bandos $^\diamondsuit$ $^\dagger$ and
 Carlos Meliveo
 $^\diamondsuit$
   \\ \vspace{1cm}
   $^\diamondsuit$Department of Theoretical Physics, University of the Basque Country,   P.O. Box 644, 48080 Bilbao,
Spain
\\
  $^\dagger$IKERBASQUE, the Basque Foundation for Science, Bilbao,
Spain
}

\abstract{We obtain the complete set of equations of motion for the interacting system of supermembrane and dynamical D=4 ${\cal N}=1$ supergravity by varying its complete superfield action  and writing the resulting  superfield equations in the special ``WZ$_{\hat{\theta}=0}$'' gauge where the supermembrane Goldstone field is set to zero ($\hat{\theta}=0$). We solve the equations for auxiliary fields and discuss the effect of dynamical generation of cosmological constant in the Einstein equation of interacting system and its renormalization due to
some regular contributions from supermembrane. These two effects (discussed in late 70th and 80th, in the bosonic perspective and in the supergravity literature) result in that, generically, the cosmological constant has different values in the branches of the spacetime separated by the supermembrane worldvolume.}

\begin{document}

\maketitle

\section{Introduction}

Soon after the covariant formulation of the ten dimensional (10D or $D=10$) Green--Schwarz superstring was found in \cite{G+S84} its  relation with partial supersymmetry breaking was appreciated \cite{Hughes:1986dn} and led to the understanding of the existence of other supersymmetric extended objects \cite{Hughes:1986fa}, which are now called super-$p$-branes \cite{AETW}. The examples presented in \cite{Hughes:1986fa} were 4D superstring ($p=1$, $D=4$) and 6D super-3-brane ($p=3$, $D=6$); $0$--branes, the massless superparticle \cite{BS81}, and massive ${\cal N}=2$ superparticles \cite{JdA+JL=82} now known under the name of D$0$-branes (Dirichlet 0-branes or Dirichlet particles), had been known before\footnote{On the other hand, to our best knowledge, the 11D massless superparticle, now called  M$0$-brane, was discussed much later, in 1996 \cite{Eric+Paul=1996}. }.

The action of the 11D supermembrane ($p=2$), now also known  under the name of M2-brane, was constructed in the famous paper \cite{BST87}.
Furthermore it was shown in \cite{BST87} that consistency of the coupling of supermembrane to supergravity background (namely, the existence of the $\kappa$--symmetry\footnote{The $\kappa$--symmetry was found as early as in 82  in the massive superparticle model  \cite{JdA+JL=82} and, independently, in the massless superparticle model in 83  \cite{Siegel83}.} in curved superspace) imposes on the background a set of superspace constraints which result in the equations of motion for the physical fields of 11D supergravity, in the same manner as the consistency conditions for the superstring coupling to 10D superfield  supergravity produces  the supergravity equations of motion \cite{Grisaru:1985fv}. The action for $D=4$ cousin of M2-brane, the ${\cal N}=1$ supersymmetric 4D supermembrane, was presented and studied in \cite{AGIT=88}.

On the other hand, the ground states of $D$ dimensional super-$p$-branes are described by supersymmetric solutions of the $D$--dimensional supergravity (SUGRA) theory \cite{Duff94,Stelle98}. These solutions are purely bosonic, although  preserve a half of the local supersymmetry characteristic for the supergravity theory. The relation of the 1/2 supersymmetry preserved by the p-brane solution of the supergravity  with the $\kappa$--symmetry of the worldvolume action for the corresponding super-$p$-brane was pointed out in \cite{Bergshoeff:1997kr}. In \cite{BdAI} it was shown that the purely bosonic limit of the supermembrane action, where the supermembrane Goldstone fermion
(the worldvolume counterpart of the Volkov--Akulov Goldstone fermion \cite{Volkov73})
is set to zero,  $\hat{\theta}^{\check{\beta}}(\xi)=0$, still preserves one half of the local supersymmetry of supergravity. This preserved $1/2$ of the target space supersymmetry is in  one--to--one correspondence with the $\kappa$--symmetry of the complete supermembrane action (the parameters of the preserved 1/2 of the supersymmetry are extracted form the generic $O(1,D-1)$ spinor with the use of $\hat{\theta}^{\check{\beta}}(\xi)=0$ limit of the supermembrane $\kappa$--symmetry projector) and is the gauge symmetry of the interacting system described by the sum of the bosonic membrane action and the action for 11D supergravity (without auxiliary fields).

The origin of this seemingly strange fact (which is clearly not restricted to the 11D supergravity--supermembrane interaction, but valid for the wide class of D dimensional super-p-branes plus SUGRA dynamical systems) was
clarified in \cite{BdAIL03,IB+JI=03} where it was shown that the sum of the purely bosonic limit of the super--$p$--brane action and the spacetime component action for supergravity can be obtained by gauge fixing from the complete superfield action of the interacting system of supergravity and super--$p$--brane given by the sum of the complete super--$p$--brane action and the superfield  action of supergravity (when the latter exists and is known).

Hence, the description of the supergravity--super-$p$--brane interacting system by the sum of the purely bosonic $p$--brane action and the spacetime component action of supergravity without auxiliary fields is called 'complete but gauge fixed' description, where ''complete'' reflects the fact that it reproduces the gauge fixed version of all the equations of the interacting system, including the  $\hat{\theta}^{\check{\beta}}(\xi)=0$ limit  of the equation for the super--$p$--brane Goldstone fermion (which is the restriction on the pull--back of the gravitino field on the super--p--brane worldvolume) \cite{BdAIL03,IB+JI=03,IB+JdA=05}\footnote{Such a completeness property does not look trivial if we remember that, after the conformal gauge is fixed in the string action, one cannot reproduce the Virasoso constraints by its variation. }. The description of such a type for the dynamical system of $D=4$ ${\cal N}=1$ supermembrane interacting with supergravity and matter multiplets was developed in \cite{Tomas+}.

Thus the interacting system of the dynamical supergravity and super--p--brane can be studied in the frame of the covariant but gauge fixed description even when the superfield action for supergravity is unknown (see \cite{IB+JdA=05} for some results on the D=11 supergravity--supermembrane interaction). However, the study of the complete superfield description, when it is known or can be obtained, in particular in $D=4$, is also interesting as a basis of possible phenomenological (model building) applications. It could also  bring some new insights on the nature of supergravity--supermembrane interaction. In particular, as we will see below, this allows to understand the role of the auxiliary fields of supergravity in some supergravity--super--$p$--brane interacting systems.

The basic examples studied in  \cite{BdAIL03,IB+JI=03}  were $D=4$ ${\cal N}=1$ supersymmetric systems of massless superparticle plus minimal supergravity \cite{BdAIL03} and superstring interacting with the minimal supergravity and tensorial supermultiplet \cite{IB+JI=03}. The study of the superfield Lagrangian description of the $D=4$ ${\cal N}=1$ supergravity--supermembrane interacting system was started in \cite{IB+CM=2011} (see \cite{IB+CM=2010} for the study of interacting system of supermembrane and $D=4$ ${\cal N}=1$ scalar multiplet), where the Wess--Zumino type approach to the   Grisaru--Siegel--Gates--Ovrut--Waldram  {\it special minimal supergravity} \cite{Grisaru:1981xm,Gates:1980az,Ovrut:1997ur} \footnote{For our best knowladge, this formulation of supergravity was discussed for the first time in \cite{Grisaru:1981xm} and \cite{Gates:1980az}, where it was noticed as a formulation with a simplified structure of the ghost sector of quantum supergravity. Ovrut and Waldram discovered it independently under the name of three form supergravity in \cite{Ovrut:1997ur} and elaborate it using an elegant mixture of the spacetime (component) and superfield formalism (much in the spirit of \cite{BW}).  } was developed and the explicit expressions for supermembrane current superfields, entering the r.h.s. of the superfield supergravity equations, were obtained.

In this paper we obtain and study the equations of motion for the   $D=4$ ${\cal N}=1$ supergravity--supermembrane interacting system. The superfield supergravity equations with supermembrane contributions are quite complex, but simplify essentially in a special gauge which we call WZ$_{\hat{\theta}=0}$ gauge. It is reached by fixing the usual Wess--Zumino (WZ) gauge for supergravity and then by using a half of the local spacetime supersymmetry  on the supermembrane worldvolume to fix the gauge where the supermembrane Goldstone fermion vanishes, $\hat{\theta}^{\check{\beta}}(\xi)=0$.
In such a gauge we solve the auxiliary field equations and show that there are three types of supermembrane contributions to the Einstein equation of the interacting system. Besides the expected singular terms with the support on the (super)membrane worldvolume $W^3$, the supermembrane produces two types of regular contributions which can be considered as contributions to the cosmological constant on the pieces of space-time separated by the supermembrane.
The first, known from the study in \cite{Ovrut:1997ur} (and before in   \cite{Aurilia:1978qs,OS80,Aurilia:1980xj,Duff:1980qv}, see \cite{IB+CM=2011} for  more references and discussion on them), is the cosmological constant generated dynamically\footnote{Although it is not proportional to the supermembrane tension and is a result of the structure of the auxiliary field sector of special minimal supergravity, this dynamically generated cosmological constant should be considered as a contribution from supermembrane as far as the requirement that supergravity should have such auxiliary field, and not the ones of the generic minimal supergravity, comes from selfconsistency of the dynamical supergravity interaction with supermembrane.}.
The second nonsingular contribution from the supermembrane changes the value of the cosmological constant on one side of the supermembrane worldvolume $W^3$ making the values of cosmological constant in two branches of spacetime $M_+^4$ and $M_4^-$ separated by $W^3$ generically different. This effect, which can be referred to as shift or renormalization of the cosmological constant by supermembrane contributions, was discussed in pure bosonic perspective in \cite{Aurilia:1978qs} and \cite{Brown:1988kg}.

Thus generically one can expect that the ground state solution of our interacting system describes the supermembrane separating two branches of AdS spaces with different values of the cosmological constant. In the purely bosonic context the solutions of such a type were studied (besides \cite{Aurilia:1978qs} and \cite{Brown:1988kg}) in  \cite{Gogberashvili:1998iu,JMMS+=2001,shellUni}. The solution for the particular case of coincident cosmological constants can be found in \cite{Ovrut:1997ur}.
In section 5 of this paper we present a preliminary discussion on possible supersymmetric solutions of the interacting system equations with supermembrane separating the asymptotically--AdS  spaces with different values of the cosmological constants.

The paper is organized as follows. We begin in Sec. 2 by a brief review of the (generic) minimal formulation of the superfield supergravity (secs. 2.1 and 2.2). In sec. 2.3 we present  the supermembrane action in minimal supergravity background, discuss  its $\kappa$--symmetry and obtain the supermembrane equations of motion in such background; these formally coincide with the supermembrane part of the superfield equations of the interacting system.

The supermembrane can propagate in a generic superspace of minimal supergravity because this allows for the existence of an invariant closed 4-form $H_4$.
The knowledge of such $H_4$  obeying $dH_4=0$ is completely sufficient for describing the supermembrane in a background superspace.  However, the existence of 3-form potential $C_3$ is crucial when one would like to find supermembrane current or to obtain the superfield equations of motion for the interacting system. As we briefly discuss in sec. 2.4, referring to  \cite{IB+CM=2011} for the detail, the requirement of the existence of the 3-form potential in  curved superspace imposes a restriction on the prepotential structure of minimal supergravity. Namely its chiral compensator becomes special chiral superfield expressed in terms of real prepotential (rather than through the complex prepotential as in the case of generic chiral superfield) \cite{IB+CM=2011}. Such a {\it special} minimal supergravity was found in \cite{Grisaru:1981xm}, discussed in \cite{Gates:1980az} and found independently and elaborated by Ovrut and Waldram \cite{Ovrut:1997ur}, under the name of 'three-form supergravity', as a supergravity formulation proper for interacting with D=4 supermembrane. In Sec. 2.4 we give a very brief review of the Wess--Zumino superfield approach to the  special minimal supergravity which was elaborated in  \cite{IB+CM=2011}; we present there the explicit expression for the variation of special minimal supergravity action which is needed to obtain the equations of the interacting system, and discuss the effect of dynamical generation of cosmological constant \cite{OS80,Ovrut:1997ur}.

In sec. 3 we present the superfield action for the supergravity---supermembrane interacting system, vary it with respect to the supergravity prepotentials (more precisely, extract the coefficients for the independent variations of the special minimal supergravity in the interacting action variation \cite{IB+CM=2011}) and find the explicit form of the  supermembrane current superfields which enter the {\it r.h.s.}-s of the superfield supergravity equations of the interacting system. Sec. 4 is devoted to the study of the spacetime component equations of the interacting system which follow from the above  mentioned superfield equations. The convenient WZ$_{\hat{\theta}=0}$ gauge is described in sec. 4.1. In sec. 4.2. the essential components of the supermembrane current superfields are calculated in this gauge and it is shown that the Rarita--Schwinger equation of the interacting system does not contain  the supermembrane contributions explicitly (although contains the spin connection which does contain supermembrane contributions).

The Einstein equation of the supergravity--supermembrane interacting system is obtained in sec. 4.3.  where the general solution of the auxiliary field equations is also given. This contains an arbitrary constant and also a contribution  proportional to the covariant version of the Heaviside  step function $\Theta (x,x_0; \hat{x})$ multiplied by the  supermembrane tension $T_2$. In sec. 4.4 we show that these result in the effect of dynamical generation of cosmological constant \cite{Aurilia:1978qs,OS80,Aurilia:1980xj,Duff:1980qv,Ovrut:1997ur} and its 'renormalization'
\cite{Aurilia:1978qs,Brown:1988kg} in such a way that the cosmological constant values in the branches of spacetime  separated by the supermembrane worldvolume are generically different. Sec. 5 contains discussion on the possible supersymmetric solution of the interacting system equations characterized by different value of cosmological constant in two branches of spacetime separated by the supermembrane worldvolume. We conclude in Sec. 6. Some details on the independent variations of the superfields of special minimal supergravity can be found in the Appendix.

\section{Superfield supergravity and supermembrane in curved D=4, ${\cal N}$=1 superspace}
\subsection{Curved superspace. Notation and conventions}
We denote local coordinates of curved $D=4$ ${\cal N}=1$ superspace $\Sigma^{(4|4)}$  by
\begin{eqnarray}\label{4Z=}
\{ {Z}^M\} \equiv   \{ x^\mu, \theta^{\underline{\breve{\alpha}}} \} \; , \qquad \mu=0,1,2,3\; ,\qquad  \underline{\breve{\alpha}}=1,2,3,4\; , \qquad
\end{eqnarray} and the bosonic and fermionic supervielbein one forms of  $\Sigma^{(4|4)}$ by
\begin{eqnarray}\label{4Ea}  E^a&=& dZ^M E_M^a(Z)\; , \quad
E^{{\alpha}}= dZ^M E_M^{{\alpha}}(Z)\; , \quad
 \bar{E}{}^{\dot\alpha}= dZ^M \bar{E}_M{}^{\dot\alpha}(Z)\; ,  \quad \\ \nonumber
 && a=0,1,2,3\,, \qquad \alpha =1,2\, , \qquad \dot{\alpha}=1,2\; .
\end{eqnarray}
Sometimes it is convenient to collect the supervielbein one forms in
\begin{eqnarray}\label{4EAM}
E^{A} = ( E^a, E^{\underline{\alpha}})  =( E^a, E^{{\alpha}}, \bar{E}{}^{{\dot\alpha}})=dZ^ME_M^A(Z)\; , \qquad
\end{eqnarray}
where ${\underline{\alpha}}=1,2,3,4$ can be understood as Majorana spinor index.
Torsion 2--forms are defined as the covariant exterior derivatives of the bosonic and fermionic supervielbein forms
\begin{eqnarray}\label{Ta:=}
T^a &:=& {\cal D}E^a=dE^a - E^b\wedge w_b{}^a= {1\over 2} E^B\wedge E^C T_{CB}{}^a
 \qquad \\
 \label{Talf:=}
T^\alpha &:=& {\cal D}E^\alpha=dE^\alpha - E^\beta\wedge w_\beta {}^\alpha= {1\over 2} E^B\wedge E^C T_{CB}{}^\alpha \; , \quad w_\beta {}^\alpha := {1\over 4} w^{ab} \sigma_{ab\beta} {}^\alpha\; , \quad \\
\label{Tdalf:=}
T^{\dot\alpha} &:=& {\cal D}E^{\dot\alpha}=dE^{\dot\alpha} - E^{\dot\beta}\wedge w_{\dot\beta} {}^{\dot\alpha}= {1\over 2} E^B\wedge E^C T_{CB}{}^{\dot\alpha} \; , \quad  w_{\dot\beta} {}^{\dot\alpha} := {1\over 4} w^{ab} \tilde{\sigma}_{ab} {}^{\dot\alpha}{}_{\dot\beta} \; , \quad
 \qquad
\end{eqnarray}
where $\omega^{ab}=-\omega^{ba}=dZ^M\omega_M^{ab}(Z)$ is the spin connection one form \footnote{$dE^A=dZ^M\wedge dE_M^A(Z)= dZ^M\wedge dZ^N\partial_N E_M^A(Z)$ and $\wedge$ is the exterior product of the differential forms, in particular,
$E^a \wedge E^b=-E^b \wedge E^a$,  $E^a \wedge E^{\underline{\beta}}=-E^{\underline{\beta}} \wedge E^a$, $E^{\underline{\alpha}} \wedge E^{\underline{\beta}}=+E^{\underline{\beta}} \wedge E^{\underline{\alpha}}$, see \cite{BdAIL03,IB+CM=2011} for more detail.},  and $\sigma^{ab}{}_{\beta} {}^\alpha=\sigma^{[a}\tilde{\sigma}^{b]}:=
{1\over 2}(\sigma^{a}\tilde{\sigma}^{b}- \sigma^{b}\tilde{\sigma}^{a})
$ and $\tilde{\sigma}{}^{ab}{}^{\dot\alpha}{}_{\dot\beta}=\tilde{\sigma}^{[a}{\sigma}^{b]} $ are antisymmetrized products of the relativistic Pauli matrices,
$\sigma^a_{\beta\dot{\alpha}}= \epsilon_{\beta\alpha} \epsilon_{\dot{\alpha}\dot{\beta}}
 \tilde{\sigma}{}^{a\dot{\beta}\alpha}$. These obey
\begin{eqnarray}\label{sasb=}
 \sigma^a\tilde{\sigma}{}^b =\eta^{ab} +{i\over 2}\epsilon^{abcd}\sigma_c\tilde{\sigma}_d\; ,\qquad \tilde{\sigma}{}^a{\sigma}^b =\eta^{ab} -{i\over 2}\epsilon^{abcd}\tilde{\sigma}_c{\sigma}_d\; ,\qquad
 \qquad
\end{eqnarray}
where $\eta^{ab}=diag (1,-1,-1,-1)$ is the Minkowski metric and $\epsilon^{abcd}=\epsilon^{[abcd]}$ is the antisymmetric
tensor with $\epsilon^{0123}=1=-\epsilon_{0123}$.

The torsion 2-forms obey the Bianchi identities
\begin{eqnarray}\label{BI=DT}
{\cal D}T^a + E^b \wedge R_b{}^a=0 \; , \qquad {\cal D}T^\alpha + E^\beta \wedge R_\beta{}^\alpha=0 \; , \qquad
 {\cal D}T^{\dot{\alpha}} + E^{\dot{\beta}} \wedge R_{\dot{\beta}} {}^{\dot{\alpha}} =0 \; , \qquad
 \qquad
\end{eqnarray}
where
\begin{eqnarray}\label{Rab}
R^{ab}= (dw-w\wedge w)^{ab}={1\over 2} E^B\wedge E^C R_{CB}{}^{ab}\; \qquad
\qquad
\end{eqnarray}
is the curvature 2-form, and  $R^{ab} =
{1\over 2}R_{\dot{\beta}} {}^{\dot{\alpha}}\tilde{\sigma}
{}^{ab\dot{\alpha}}_{\dot{
\beta}} -{1\over 2}R_{{\beta}} {}^{{\alpha}}{\sigma}
{}^{ab}{}_{\alpha}{}^{\beta} $ provides its decomposition on the anti-self-dual and self-dual parts.

\subsection{Superfield supergravity action, superspace constraints and equations of motion}

The superfield action of the minimal off-shell formulation of $D=4$, ${\cal N}=1$  supergravity \cite{WZ78}
\begin{eqnarray}\label{SGact} S_{SG} = \int d^8Z \;
E :=  \int d^4 x \tilde{d}^4\theta \; sdet(E_M^A) \; , \qquad
\end{eqnarray}
is given by the superdeterminant (or Berezinian) of the matrix of supervielbein coefficients,  $E_M^A(Z)$ in (\ref{4EAM}), which obey the set of supergravity constraints.
These can be collected together with their consequences in the following expressions for the superspace torsion 2-forms (\ref{Ta:=}), (\ref{Talf:=}) (see \cite{BdAIL03} and refs. therein)
\begin{eqnarray}\label{4WTa=}T^a &=&- 2i\sigma^a_{\alpha\dot{\alpha}} E^\alpha \wedge \bar{E}^{\dot{\alpha}} - {1\over 8} E^b \wedge E^c
\varepsilon^a{}_{bcd} G^d \; ,\;   \\ \label{4WTal=} T^{\alpha} &:=& (T^{\dot{\alpha}})^*= {i\over 8} E^c \wedge E^{\beta}
(\sigma_c\tilde{\sigma}_d)_{\beta} {}^{\alpha} G^d   -{i\over 8} E^c
\wedge \bar{E}^{\dot{\beta}} \epsilon^{\alpha\beta}\sigma_{c\beta\dot{\beta}}R +
 {1\over 2} E^c \wedge E^b \; T_{bc}{}^{\alpha}\; .\;
\end{eqnarray}
The {\it  main superfields}, real vector $G_a=(G_a)^*$ and complex scalar $R=(\bar{R})^*$, entering (\ref{4WTa=}) and (\ref{4WTal=}),  obey
\begin{eqnarray} \label{chR} & {\cal D}_\alpha \bar{R}=0\;
, \qquad \bar{{\cal D}}_{\dot{\alpha}} {R}=0\; ,
\\
\label{DG=DR} & \bar{{\cal
D}}^{\dot{\alpha}}G_{{\alpha}\dot{\alpha}}= - {\cal D}_{\alpha} R \; , \qquad {{\cal
D}}^{{\alpha}}G_{{\alpha}\dot{\alpha}}= - \bar{{\cal D}}_{\dot{\alpha}} \bar{R} \; .
 \qquad \end{eqnarray}
These relations can be obtained by studying the Bianchi identities (\ref{BI=DT}), which also allow to find the expression for  superfield generalization of the gravitino field strength, $T_{bc}{}^{\alpha}(Z)$,
\begin{eqnarray}\label{Tabg} T_{{\alpha}\dot{\alpha}\; \beta \dot{\beta }\; {\gamma}} &
\equiv   \sigma^a_{{\alpha}\dot{\alpha}} \sigma^b_{\beta \dot{\beta }} \epsilon_{{\gamma}{\delta}}
T_{ab}{}^{{\delta}}=  -{1\over 8}  \epsilon_{{\alpha}{\beta}} {\bar{{\cal D}}}_{(\dot{\alpha}|} G_{\gamma
|\dot{\beta})} - {1\over 8} \epsilon_{\dot{\alpha}\dot{\beta}}[W_{\alpha \beta\gamma} -
2\epsilon_{\gamma (\alpha}{\cal D}_{\beta)} R]
 \; , \end{eqnarray}
involving one more main superfield, $W_{\alpha \beta\gamma}=W_{(\alpha \beta\gamma )}=: (\bar{W}_{\dot{\alpha}\dot{\beta}\dot{\gamma}})^*$. This  obeys
 \begin{eqnarray}
 \label{chW} & \bar{{\cal D}}_{\dot{\alpha}} W^{\alpha\beta\gamma}= 0\; , \qquad
{{\cal D}}_{{\alpha}}\bar{W}^{\dot{\alpha}\dot{\beta}\dot{\gamma}}= 0\;,
 \\ \label{DW=DG} & {{\cal D}}_{{\gamma}}W^{{\alpha}{\beta}{\gamma}}= \bar{{\cal D}}_{\dot{\gamma}} {{\cal
D}}^{({\alpha}}G^{{\beta})\dot{\gamma}} \; . \qquad
\end{eqnarray}

Studying the Bianchi identities with the constraints   (\ref{4WTa=}), (\ref{4WTal=})  one also finds that the superfield generalization of  the left hand side ({\it l.h.s.}) of the supergravity Rarita--Schwinger equation reads
\begin{eqnarray}\label{SGRS=off}
\epsilon^{abcd}T_{bc}{}^{\alpha}\sigma_{d\alpha\dot{\alpha}} ={i\over 8}
\tilde{\sigma}^{a\dot{\beta}\beta} \bar{{\cal D}}_{(\dot{\beta}|} G_{\beta|\dot{\alpha})} + {3i\over 8}
{\sigma}^a_{\beta \dot{\alpha}} {\cal D}^{\beta}R \; ,
\end{eqnarray}
and the superfield generalization of the Ricci tensor is
\begin{eqnarray} \label{RRicci}
R_{bc}{}^{ac}& = {1\over 32} ({{\cal D}}^{{\beta}} \bar{{\cal D}}^{(\dot{\alpha}|}
G^{{\alpha}|\dot{\beta})} - \bar{{\cal D}}^{\dot{\beta}} {{\cal D}}^{({\beta}}G^{{\alpha})\dot{\alpha}})
\sigma^a_{\alpha\dot{\alpha}}\sigma_{b\beta\dot{\beta}} - {3\over 64}
(\bar{{\cal D}}\bar{{\cal D}}\bar{R} + {{\cal D}}{{\cal D}}{R}- 4 R\bar{R})\delta_b^a\; .
\end{eqnarray}
This suggests that superfield supergravity equation should have the form
 \begin{eqnarray}\label{SGeqmG}
 && G_a =0 \; , \qquad
\\ \label{SGeqmR}
 && R=0 \; ,  \qquad \bar{R}=0\; . \qquad
\end{eqnarray}
See \cite{BdAIL03,IB+CM=2011} for more detail on the superfield description of minimal supergravity in the present notation.

Eqs. (\ref{SGeqmG}) and (\ref{SGeqmR}) can be obtained by varying the action (\ref{SGact}) with respect to supervielbein obeying the supergravity constraints
(\ref{4WTa=}), (\ref{4WTal=}) \cite{WZ78}. Such admissible variations are expressed through a vector parameter
$\delta {\cal H}^a$ and complex scalar parameter $\delta {\cal U}= (\delta \bar{{\cal U}})^*$ which enter the variation of the supervielbein and spin connection under the symbol of the chiral projector  $({\cal D}^\alpha{\cal D}_\alpha -\bar{R})$ (see \cite{BdAIL03,IB+CM=2011} for more detail). They  correspond to the variations of the so-called {\it prepotentials}, unconstrained superfields which appear in the  general solution of the supergravity constraints. The minimal supergravity constraints are solved in terms of the axial vector superfield ${\cal H}^\mu$ \cite{OS78} and chiral compensator $\Phi$ \cite{Siegel:1978nn}. This latter obeys $\bar{{\cal D}}_{\dot{\alpha}}\Phi=0$ and, hence, can be expressed as $\Phi =(\bar{{\cal D}}_{\dot{\alpha}}\bar{{\cal D}}{}^{\dot{\alpha}}-R){\cal U}$ with a complex unconstrained superfield
${\cal U}$.

Thus the set of minimal supergravity prepotentials includes ${\cal H}^\mu$, ${\cal U}$ and  $\bar{{\cal U}}=({{\cal U}})^*$ which are in one to one correspondence with the set of three independent variations $\delta {\cal H}^a$, $\delta {\cal U}$ and  $\delta \bar{{\cal U}}=(\delta  {{\cal U}})^*$ of the Wess--Zumino approach to supergravity \cite{WZ77,WZ78} producing the three superfield equations (\ref{SGeqmG}) and  (\ref{SGeqmR}). In short, as it had been known already from \cite{WZ78},
 \begin{eqnarray}\label{vSGsf} \delta S_{SG} =  \int
d^8Z E\;  \left[{1\over 6} G_a \; \delta H^a -  2 R\; \delta \bar{{\cal U}} -2 \bar{R}\; \delta {\cal U}\right] \; . \quad
\end{eqnarray}

\subsection{Supermembrane action in minimal supergravity background}

As it is well known, the supermembrane action \cite{BST87,AGIT=88} is given by the sum of the Dirac--Nambu--Goto and the Wess--Zumino term,
  \begin{eqnarray}\label{Sp=2:=}
  S_{p=2}= {1\over 2}\int d^3 \xi \sqrt{g} - \int\limits_{W^3} \hat{C}_3 = -{1\over 6} \int\limits_{W^3} *\hat{E}_a\wedge \hat{E}{}^a - \int\limits_{W^3} \hat{C}_3\; . \qquad
\end{eqnarray}
The former is given by the volume of $W^3$ defined as  integral of the  determinant of the induced metric, $g=det(g_{mn})$,
\begin{eqnarray}\label{g=EE}
g_{mn}= \hat{E}_m^{a}\eta_{ab} \hat{E}_n^{b}\; , \qquad
\hat{E}_m^{a}:= \partial_m \hat{Z}^M(\xi) E_M^a(\hat{Z})\; .  \qquad
\end{eqnarray}
Here $\xi^m= (\tau,\sigma^1, \sigma^2)$ are local coordinates on $W^3$ and $\hat{Z}^M(\xi)$ are coordinates functions which determine the embedding of $W^3$ as a surface in target superspace $\Sigma^{(4|4)}$,
\begin{eqnarray}\label{W3inS44}
W^3\; \subset  \Sigma^{(4|4)}\; : \qquad Z^M= \hat{Z}{}^{{ {M}}}(\xi)= (\hat{x}{}^{\mu}(\xi)\, ,
\hat{\theta}^{\breve{\alpha}}(\xi))\; . \;
\end{eqnarray}
In the second equality  of (\ref{Sp=2:=}) the Dirac--Nambu--Goto term is written  as an integral of the wedge product of the pull--back of the $\Sigma^{(4|4)}$ bosonic supervielbein form $E^a$ to $W^3$,
\begin{eqnarray}\label{hEa=dxiE}
\hat{E}{}^a= d\xi^m \hat{E}_m^{a} = d \hat{Z}^M(\xi) E_M^a(\hat{Z})\; ,  \qquad
\end{eqnarray}
and of its Hodge dual two form defined with the use of the induced metric (\ref{g=EE}) and its inverse $g^{mn}$,
\begin{eqnarray}\label{*Ea:=}
*\hat{E}^a:= {1\over 2}d\xi^m\wedge d\xi^n\sqrt{g}\epsilon_{mnk}g^{kl}\hat{E}_l^a \; . \qquad \end{eqnarray}

The second, Wess--Zumino term of the supermembrane action (\ref{Sp=2:=}) describes the  supermembrane coupling to a 3--form gauge potential $C_3$ defined on $\Sigma^{(4|4)}$,
\begin{eqnarray} \label{C3:=}
 C_3= {1\over 3!} dZ^M\wedge dZ^N\wedge dZ^K C_{KNM}(Z)= {1\over 3!} E^C\wedge E^B\wedge E^A C_{ABC}(Z)\;  .  \qquad
\end{eqnarray}
Thus, to write a supermembrane action, one has to construct the 3--form gauge potential $C_3$ in the target superspace $\Sigma^{(4|4)}$ and take its pull--back to the supermembrane worldvolume
\begin{eqnarray} \label{C3:=}
 \hat{C}_3 &=& {1\over 3!} d\hat{Z}^M\wedge d\hat{Z}^N\wedge d\hat{Z}^K C_{KNM}(\hat{Z}(\xi))=
 {1\over 3!} \hat{E}^C\wedge \hat{E}^B\wedge \hat{E}^A C_{ABC}(\hat{Z})= \qquad \nonumber \\ &=&  {1\over 3!} d\xi^m\wedge d\xi^n\wedge d\xi^k \hat{C}_{knm} =
   d^3\xi \epsilon^{mnk}  \hat{C}_{knm}\;  .  \qquad
\end{eqnarray}

Actually, to study {\it supermembrane in supergravity background}, it is sufficient to know the field strength of the above 3--form potential, $H_4=dC_3$. This should be closed, $dH_4=0$, and supersymmetric invariant 4--form.
In flat superspace such a form exists and represents a nontrivial Chevalley--Eilenberg cohomology  of the ${\cal N}=1$ supersymmetry algebra \cite{JdA+PKT89,Jose+Paul=PRL89,AI95}.

The minimal supergravity superspace  allows for existence of two closed 4-forms
\begin{eqnarray} \label{H4L}  & H_{4L} =  - {i\over 4} E^b\wedge E^a \wedge E^\alpha \wedge E^\beta \sigma_{ab\; \alpha\beta} -   {1\over 128} E^{d} \wedge E^c \wedge E^b \wedge E^a \epsilon_{abcd} R \, , \qquad
 dH_{4L}=0 \, ,
\end{eqnarray}
and its complex conjugate $H_{4R}=(H_{4L})^*$ (see \cite{IB+CM=2011}).
Its real part,
\begin{eqnarray} \label{H4=HL+HR}  H_4&:=&dC_3= {1\over 4!} E^{A_4}\wedge ... \wedge E^{A_1}
H_{A_1\ldots A_4}(Z)= H_{4L}+H_{4R}\;  ,   \qquad
\end{eqnarray}
is also closed  and  provides  the  4--form field strength associated to the Wess--Zumino (WZ) term of the supermembrane action  in the minimal supergravity background  \cite{Ovrut:1997ur}, $\int_{W^3} C_3$ in (\ref{Sp=2:=}). Indeed, the WZ term can also be defined as an integral of the closed 4 form $H_4$,  related to $C_3$ by $H_4=dC_3$, over some four dimensional space ${W}^4$ the boundary of which is given by the supermembrane worldvolume $W^3$,
\begin{eqnarray} \label{WZ=intH4}
\int\limits_{W^3}{C}_3= \int\limits_{W^4\;:\; \partial W^4=W^3}{H}_4
\;  .    \qquad
\end{eqnarray}
The condition that the form $H_4$ is closed, $dH_4=0$, guaranties that the integral $\int_{{W}^4}H_4$ is independent on the choice of ${W}^4$ and, thus, is related to the supermembrane worldvolume $W^3$.

The fact that  the knowledge of $H_4$ is completely sufficient for studying the properties of closed supermembrane in a supergravity background is related to that in this case the only dynamical variables are the supermembrane coordinate functions $\hat{Z}^M(\xi)$, that the action is written in term of pull--back of differential forms to $W^3$, and that the variation of the differential form with respect to the coordinates can be calculated with the use of the Lie derivative formula, in particular $\delta_{\delta Z}C_3= i_{\delta Z} H_4 + di_{\delta Z}  C_3=
{1\over 3!} E^{A_4}\wedge ... \wedge E^{A_2} \; \delta Z^{M} \, E_M^{A_1}\,
H_{A_1\ldots A_4}(Z) + d({1/2} E^{C}\wedge E^{B} \; \delta Z^{M} \, E_M^{A}\,
C_{ABC}(Z))$.  \footnote{
For closed supermembrane $\partial W^3= \emptyset$ so that $\int_{W^3}d\alpha_2 =\int_{\partial W^3}\alpha_2 = 0$  for any 2-form $\alpha_2$, including for $\alpha_2=i_{\delta Z} C_3$.}

The supermembrane equations of motion in the minimal supergravity background, which are obtained by varying the action (\ref{Sp=2:=}) with respect to the coordinate functions $\delta \hat{Z}^M(\xi)$,
\begin{eqnarray} \label{vSp2:=}
\delta S_{p=2}= \int_{W^3} \left({1\over 2} {\cal M}_{3a} E_M{}^a (\hat{Z})+  i \Psi_{3\alpha} E_M{}^\alpha (\hat{Z}) + i \Psi_{3\dot\alpha} E_M{}^{\dot\alpha} (\hat{Z})  \right)
\, \delta \hat{Z}^M(\xi)
\;  ,    \qquad
\end{eqnarray}
read
\begin{eqnarray}\label{SmEqm=b}
 {\cal M}_{3\, a}&:=& {\cal D}* \hat{E}_a + i \hat{E}{}^b\wedge \hat{E}{}^{\alpha}\wedge \hat{{E}}{}^{{\beta}}{\sigma}_{ab{\beta}{\alpha}} -  i  \hat{E}{}^b\wedge \hat{\bar{E}}{}^{\dot{\alpha}} \wedge \hat{\bar{E}}{}^{\dot{\beta}}\tilde{\sigma}_{ab\dot{\beta}\dot{\alpha}} - \qquad \nonumber \\ &&
-{1\over 8} \hat{E}{}^b\wedge \hat{E}{}^c\wedge \hat{E}{}^d\epsilon_{abcd} (R+\bar{R})
=0
\;   \qquad
\end{eqnarray}
and
\begin{eqnarray}\label{SmEqm=f}
 \bar{\Psi}{}_{3\dot{\alpha}}:= *\hat{E}_a\wedge \left(\hat{E}{}^\alpha {\sigma}{}^a_{{\alpha}\dot{\alpha}}-
 (\tilde{\bar{\gamma}} {\sigma}{}^a)_{\dot{\alpha}\dot{\beta}} \hat{\bar{E}}{}^{\dot{\beta}}\right)=0 \; , \qquad
 \\
 \label{SmEqm=bf}
 {\Psi}{}_{3{\alpha}}:= *\hat{E}_a\wedge \left( {\sigma}{}^a_{{\alpha}\dot{\alpha}} \hat{\bar{E}}{}^{\dot{\alpha}}+
  \hat{{E}}{}^{{\beta}} ( {\sigma}{}^a\tilde{\bar{\gamma}})_{{\alpha}{\beta}}\right)=0 \; , \qquad
  \end{eqnarray}
where the matrix $\bar{\gamma}_{\beta\dot{\alpha}}$ is defined by
  \begin{eqnarray}\label{tg:=}
  \bar{\gamma}_{\beta\dot{\alpha}}= \epsilon_{\beta\alpha}\epsilon_{\dot{\alpha}\dot{\beta}}\tilde{\bar{\gamma}}{}^{\dot{\beta}{\alpha}}= {i\over 3!\sqrt{g}} \sigma^a_{\beta\dot{\alpha}}\epsilon_{abcd} \epsilon^{mnk} \hat{E}{}_m^b\hat{E}{}_n^c\hat{E}{}_k^d =- (
  \bar{\gamma}_{{\alpha}\dot{\beta}})^*\;  \qquad
\end{eqnarray}
and obeys
\begin{eqnarray}
  \label{tgtg=I} \bar{\gamma}_{\beta\dot{\alpha}}\tilde{\bar{\gamma}}{}^{\dot{\alpha}{\alpha}}=  \delta_{\beta}{}^{{\alpha}}\; , \qquad\tilde{\bar{\gamma}}{}^{\dot{\alpha}{\alpha}}  \bar{\gamma}_{\alpha\dot{\beta}}=  \delta^{\dot{\alpha}}{}_{\dot{\beta}}\; . \qquad
\end{eqnarray}
Some identities involving the above matrix are
\begin{eqnarray}
  \label{tgs=stg+} \bar{\gamma}\tilde{\sigma}{}^a = -  {\sigma}{}^a \tilde{\bar{\gamma}} +
 {i\over 3!\sqrt{g}} \epsilon_{abcd} \epsilon^{mnk} \hat{E}{}_m^b\hat{E}{}_n^c\hat{E}{}_k^d
\; , \qquad
\\  \label{Etgsa=}
*\hat{E}_a \bar{\gamma}\tilde{\sigma}{}^a\bar{\gamma}= *\hat{E}_a {\sigma}{}^a
\; , \qquad *\hat{E}_a \bar{\gamma}\tilde{\sigma}{}^a=- *\hat{E}_a {\sigma}{}^a \tilde{\bar{\gamma}}
\; , \qquad
\\  \label{EEs=*Esg}
{1\over 2} \;
 \hat{E}{}^b\wedge \hat{E}{}^a\wedge \hat{{E}}{}^{{\beta}}\; {\sigma}_{ab{\beta}{\alpha}} = \; *\hat{E}_a \wedge \hat{{E}}{}^{{\beta}}({\sigma}^a\tilde{\bar{\gamma}})_{{\beta}{\alpha}} \; , \qquad
\\  \label{EEts=*Etsg} {1\over 2}
 \hat{E}{}^b\wedge \hat{E}{}^a\wedge \hat{\bar{E}}{}^{\dot{\beta}}\tilde{\sigma}_{ab\dot{\beta}\dot{\alpha}}=
- *\hat{E}_a \wedge \hat{\bar{E}}{}^{\dot{\beta}} (\tilde{\sigma}{}^a\bar{\gamma})_{\dot{\beta}\dot{\alpha}}
\; . \qquad
\end{eqnarray}
They are useful, in particular, to show that  the fermionic equations of motion obey the Noether identity $\bar{\Psi}{}_{3\dot{\alpha}}= {\Psi}{}_{3}^{\alpha}{\bar{\gamma}}_{{\alpha}\dot{\alpha}} $ reflecting the local fermionic $\kappa$--symmetry
\begin{eqnarray}
  \label{kappaZ=}
\delta_\kappa \hat{Z}^M = \kappa^\alpha (\xi)(E_\alpha^M(\hat{Z})+ \bar{\gamma}_{\alpha\dot{\alpha}}\epsilon^{\dot{\alpha}\dot{\beta}}E_{\dot{\beta}}^M(\hat{Z})) \; \qquad
\end{eqnarray}
with the local fermionic ``parameter''  $\kappa^\alpha (\xi) =({\bar{\kappa}}{}^{\dot{\alpha}})^*$ obeying
\begin{eqnarray}
  \label{kappa=*}
\kappa^\alpha (\xi)  = -
{\bar{\kappa}}_{\dot{\alpha}}(\xi) \tilde{\bar{\gamma}}{}^{\dot{\alpha}{\alpha}} \; \qquad  \Leftrightarrow \; \qquad
{\bar{\kappa}}_{\dot{\alpha}} (\xi) = - \kappa^\alpha (\xi)  {\bar{\gamma}}_{{\alpha}\dot{\alpha}}\; .
\end{eqnarray}
The relation of the supermembrane $\kappa$--symmetry in curved superspace with the minimal supergravity constraints was discussed  in \cite{Ovrut:1997ur}. The flat superspace limit of our equations reproduces the equations of the seminal paper \cite{AGIT=88}.

\subsection{3--form potential in the minimal supergravity  superspace. Special minimal supergravity  }


Thus, as we have seen in the previous subsection,  to find the equations of motion of supermembrane in supergravity background as well as to study its symmetries it is sufficient to know the closed 4-form $H_4=dC_3$ in the background superspace.

However, to calculate the supermembrane current(s) describing the supermembrane contribution(s) to the supergravity (super)field equations, one needs to vary the Wess--Zumino term $\int \hat{C}_3$ of the supermembrane action with respect to the supergravity (super)fields. Thus one arrives at a separate problem of  finding the variation
\begin{eqnarray} \label{vC3:=}
 \delta C_3= {1\over 3!} E^C\wedge E^B\wedge E^A \beta_{ABC}(\delta)\;    \qquad
\end{eqnarray}
such that $d\delta C_3=\delta H_4$ reproduces the variation of $H_4$ from (\ref{H4=HL+HR}), (\ref{H4L}),
written in terms of the basic supergravity variations (we refer to \cite{IB+CM=2011} for the explicit expression of $\delta H_4$).

Studying such a technical problem we have  found \cite{IB+CM=2011} that it imposes a restriction on the independent variations of the supergravity prepotentials, or equivalently, on the independent parameters of the admissible supervielbein variations,  thus transforming the generic minimal supergravity into a {\it special minimal supergravity}.
This off--shell supergravity formulation had been described for the first time in \cite{Grisaru:1981xm}, further discussed in \cite{Gates:1980az} (see also latter \cite{Kuzenko+05})  and elaborated in  \cite{Ovrut:1997ur} using the elegant combination of superfield results and the component 'tensor calculus' approach on the line of \cite{BW}.

In \cite{IB+CM=2011}  we described this {\it special minimal supergravity} in the complete Wess--Zumino superfield formalism.  Referring to that paper for technical details, we only notice that the existence of the 3-form potential imposes a restriction on the prepotential structure of minimal supergravity which in our approach manifests itself in that the basic complex variations ${\delta U}$ and   ${\delta \bar{U}}=({\delta U})^*$ are expressed in terms of one real variation $\delta V$, essentially
\begin{eqnarray}
 \label{vcU=vV+}
\delta {\cal U} = {i\over 12}\delta {V} \; ,   \qquad \delta \bar{{\cal U}} = - {i\over 12}\delta {V} \;  . \qquad
\end{eqnarray}
As a result, the variation of the special minimal supergravity action  is essentially (see \cite{IB+CM=2011} and the Appendix)
\begin{eqnarray}\label{vSGsf=s}  \delta S_{SG} &=&  {1\over 6} \int
d^8Z E\;  \left[ G_a \; \delta H^a + (R-\bar{R}) i\delta {V}  \right]
 \; . \qquad
\end{eqnarray}
Hence the set of superfield equations of special minimal supergravity still includes
 the vector superfield equation (\ref{SGeqmG}),
\begin{eqnarray}\label{SGeqmG=0}
G_a=0\; , \qquad
\end{eqnarray}
but instead of the complex scalar superfield equations (\ref{SGeqmR}), valid in the case of generic minimal supergravity, in the case of special minimal supergravity we have only the real scalar equation
\begin{eqnarray}\label{SGeqmR+*=0}
R- \bar{R}=0\; . \qquad
\end{eqnarray}
Clearly, due to chirality of $R$, $\;\bar{{\cal D}}_{\dot\alpha} {R}=0$,  and anti-chirality of $\bar{R}$, $\; {\cal D}_\alpha \bar{R}=0$, the above Eq. (\ref{SGeqmR+*=0}) also implies that $d(R+\bar{R})=0$ so that on the mass shell
the complex superfield $R$ is actually equal to a {\it real}  constant,
\begin{eqnarray}\label{bR=4ic}
R=4c \; , \qquad \bar{R}=4c\; , \qquad c=const=c^*\; .
\end{eqnarray}
Using (\ref{RRicci}), one finds that the superfield equation (\ref{SGeqmR+*=0}) results in Einstein equation with cosmological constant
\begin{eqnarray}\label{RRici=c}
R_{bc}{}^{ac}= 3c^2 \delta_b{}^a\; .
\end{eqnarray}
The value of the cosmological constant is proportional to the square of the above arbitrary constant $c$, which has appeared as an integration constant, so that the special minimal supergravity is characterized by a {\it cosmological constant generated dynamically}.

The above mechanism of the dynamical generation of cosmological constant in special minimal supergravity is the same as was observed by  Ogievetski and Sokatchev \cite{OS80} in their theory of axial vector superfield. In the language of component spacetime approach to supergravity the dynamical generation of cosmological constant in the special minimal supergravity was described in \cite{Ovrut:1997ur} and before it, in purely bosonic perspective in \cite{Aurilia:1978qs,Duff:1980qv,Brown:1988kg} and in the context of spontaneously broken $N=8$ supergravity in \cite{Aurilia:1980xj}.
We refer to \cite{IB+CM=2011} for more references and discussion.

\section{Superspace action and superfield equations of motion for the interacting system of dynamical supergravity and supermembrane }
\label{SGacSGeq}

The action for interacting system of dynamical supergravity and supermembrane reads
\begin{eqnarray}\label{Sint=SG+Sp2}
 S&=& S_{SG}+ T_2 S_{p=2}= \int d^8 Z E(Z) +
 {T_2\over 2}\int d^3 \xi \sqrt{g} - T_2 \int\limits_{W^3} \hat{C}_3\; ,
\end{eqnarray}
where $S_{p=2} $ is the same as in Eq. (\ref{Sp=2:=}), the supervielbein (\ref{4EAM}) and the 3-form potential (\ref{C3:=}) are assumed to be restricted by the minimal supergravity constraints (\ref{4WTa=}), (\ref{4WTal=}), (\ref{H4=HL+HR}), (\ref{H4L}). Furthermore, as we have discussed in previous sections (and in more details in \cite{IB+CM=2011}; see also Appendix), the existence of the 3-form potential imposes the restrictions (\ref{vcU=vV+}) on the prepotentials of minimal supergravity or equivalently on the basic supergravity variations.

As a result, the  superfield equations which appear as a result of variation of the interacting action (\ref{Sint=SG+Sp2}) read
 \cite{IB+CM=2011}
\begin{eqnarray}\label{Ga=Ja}
 G_a= T_2J_a\; ,
\end{eqnarray}
and
\begin{eqnarray}\label{R-bR=cX}
R-\bar{R}= -i T_2{\cal X}
\end{eqnarray}
where $J_a$ and ${\cal X}=({\cal X})^*$ are supermembrane scalar superfields. Roughly speaking, they are obtained as a result of varying the supermembrane action with respect to the prepotentials of the special minimal supergravity, this is to say  as $\delta S_{p=2} /\delta H^a$ and $\delta S_{p=2} /\delta V$ \cite{IB+CM=2011}, and have the form
\begin{eqnarray}\label{Ja=NG+WZ}
J_a&=&
\int\limits_{W^3} {3\over \hat{E}} \hat{E}{}^b \wedge \hat{E}{}^\alpha \wedge \hat{E}{}^\beta\; {\sigma}_{ab \alpha\beta}  \delta^8 (Z-\hat{Z}) - \qquad \nonumber
\\ && -
\int\limits_{W^3} {3i\over \hat{E}} \left( *\hat{E}_a \wedge \hat{E}{}^\alpha + {i\over 2}   \hat{E}{}^b \wedge \hat{E}{}^c \wedge \hat{\bar{E}}_{\dot{\beta}}\epsilon_{abcd} \tilde{\sigma}^{d\dot{\beta}\alpha}\right) {\cal D}_\alpha \delta^8 (Z-\hat{Z})
+c.c -
\nonumber
\\
&& -  \int\limits_{W^3}  {i\over 8\hat{E}}   \, \hat{E}{}^b \wedge \hat{E}{}^c \wedge \hat{E}{}^d \, \epsilon_{abcd} \left( {\cal D}{\cal D}- {1\over 2}\bar{R} \right) \delta^8 (Z-\hat{Z}) + c.c.  + \qquad \nonumber
\\ && + \int\limits_{W^3}  {1\over 4\hat{E}} *\hat{E}_b \wedge \hat{E}{}^b  \, G_a\,  \delta^8 (Z-\hat{Z})  -  \nonumber  \qquad
\\
&& -\int\limits_{W^3}  {1\over 4\hat{E}}\; *\hat{E}_c \wedge \hat{E}{}^b \tilde{\sigma}^{d\dot{\alpha}\alpha} \left( 3\delta_a^c \delta_b^d-  \delta_a^d \delta_b^c \right)[{\cal D}_\alpha , \bar{\cal D}_{\dot{\alpha}}] \delta^8 (Z-\hat{Z})  \; , \quad
\end{eqnarray}
and
\begin{eqnarray}\label{cX=}
{\cal X}& =&  {6i\over {E}}  \int\limits_{W^3} \hat{E}^a \wedge \hat{E}{}^\alpha  \wedge \hat{\bar{E}}{}^{\dot{\alpha}}\, {\sigma}^a_{\alpha\dot{\alpha}}\; \delta^8 (Z-\hat{Z}) -
\nonumber  \qquad
\\  && - {3\over 2}\int\limits_{W^3} { \hat{E}{}^b \wedge \hat{E}{}^a \wedge \hat{E}{}^\alpha \over \hat{E}}   \; {\sigma}_{ab \alpha}{}^{{\beta}} {\cal D}_{\beta} \delta^8 (Z-\hat{Z})
+c.c + \nonumber  \;
\\ && +  \int\limits_{W^3}  {   \, \hat{E}{}^b \wedge \hat{E}{}^c \wedge \hat{E}{}^d \over 8 \hat{E}}\, \epsilon_{abcd} \tilde{\sigma}^{a\dot{\alpha}\alpha} [{\cal D}_\alpha , \bar{\cal D}_{\dot{\alpha}}] \delta^8 (Z-\hat{Z})+\;
\nonumber  \;
\\ && + i\int\limits_{W^3}  {*\hat{E}_a \wedge \hat{E}{}^a \over 4\hat{E}}
\left( {\cal D}{\cal D}- \bar{R} \right) \delta^8 (Z-\hat{Z}) + c.c. + \;  \qquad \nonumber \;
\\&& +\int\limits_{W^3}  {1 \over 4\hat{E}} \hat{E}{}^b \wedge \hat{E}{}^c \wedge \hat{E}{}^d \epsilon_{abcd} G{}^{a}\; \delta^8 (Z-\hat{Z})\; . \qquad
\end{eqnarray}

Notice that, as a consequence of (\ref{DG=DR}), the supermembrane current superfields obey
\begin{eqnarray}\label{DJ=-iDX}
\bar{{\cal D}}{}^{\dot{\alpha}} J_{\alpha\dot{\alpha}} = i {{\cal D}}_{{\alpha}} {\cal X}
\; , \qquad
{{\cal D}}{}^{{\alpha}} J_{\alpha\dot{\alpha}} = -i \bar{{\cal D}}_{\dot{\alpha}}{\cal X}\; .
\end{eqnarray}
Although at first glance these relations look different from any of listed in \cite{Seiberg+2009,Kuzenko:2010}, they can be reduced to the Ferrara--Zumino multiplet \cite{FerraraZumino74} if one takes into account Eq. (\ref{R-bR=cX}). Indeed, this states that the real superfield ${\cal X}$ in the {\it r.h.s.} of Eq. (\ref{DJ=-iDX}) is the sum of chiral superfield (equal to $iR$) and its complex conjugate, so that only the first (second) one contributes to the {\it r.h.s.} of the first (second) equation in (\ref{DJ=-iDX}).

The explicit form of supercurrent superfields (\ref{Ja=NG+WZ}), (\ref{cX=}) was presented in  \cite{IB+CM=2011} were one can find more detail on their derivation. The main aim of this paper is to  study the spacetime (component) equations which are encoded in the superfield equations (\ref{Ga=Ja}) and (\ref{R-bR=cX}) with the supermembrane contributions (\ref{Ja=NG+WZ}) and  (\ref{cX=}).

\section{Spacetime component equations of the $D=4$ ${\cal N}=1$ supergravity--supermembrane interacting system}

\subsection{Wess--Zumino gauge plus partial gauge fixing of the  local spacetime supersymmetry (WZ$_{\hat{\theta}=0}$ gauge)}

The structure of the current superfields (\ref{Ja=NG+WZ}), (\ref{cX=}) is quite complicated. So is the structure of their components. To simplify the supercurrent  components which contribute to the equations of physical, spacetime component fields, we use the general coordinate invariance to fix the Wess--Zumino (WZ) gauge  on supergravity superfields,
\begin{eqnarray}  \label{WZgauge=f} i_{\underline{\theta}}  E^{
{\alpha}}&:=& \theta^{\breve{\underline{\alpha}}} E_{\breve{\underline{\alpha}}}{}^{\alpha}
= \theta^{ {\alpha}} \; , \qquad i_{\underline{\theta}}  E^{\dot{\alpha}}:=\theta^{\breve{\underline{\alpha}}} E_{\breve{\underline{\alpha}}}{}^{\dot\alpha}
=  \bar{\theta}{}^{ \dot{\alpha}} \; , \qquad \\  \label{thal=} &&  \theta^{ {\alpha}}:= \theta^{\breve{\underline{\beta}}} \delta_{\breve{\underline{\beta}}}^{\,
 {\alpha}}\; , \qquad  \bar{\theta}{}^{ \dot{\alpha}}:= \theta^{\breve{\underline{\beta}}} \delta_{\breve{\underline{\beta}}}^{\,\dot{\alpha}}\; , \qquad \\ \label{WZgauge=b}
i_{\underline{\theta}} E^{\underline{a}}& :=& \theta^{\breve{\underline{\alpha}}} E_{\breve{\underline{\alpha}}}{}^{\underline{a}}=0\; , \quad \\
 \label{WZgauge=w}i_\theta w^{\underline{a}\underline{b}}&:=&  \theta^{\breve{\underline{\beta}}} w_{\breve{\underline{\beta}}}^{\underline{a}\underline{b}}= 0\;  \qquad
\end{eqnarray}
(see \cite{BdAIL03} for references and more detail) and the (pull--back to $W^3$ of the) local spacetime supersymmetry to set to zero the fermionic Goldstone field of the supermembrane,
\begin{eqnarray}\label{thGauge}
\hat{\theta}{}^{\underline{\alpha}}(\xi) =0\qquad \Leftrightarrow \qquad \hat{\theta}{}^{{\alpha}}(\xi) =0\; , \qquad \hat{\bar{\theta}}{}^{\dot{\alpha}}(\xi) =0
\; .  \qquad
\end{eqnarray}
A detailed discussion on this  ''WZ$_{\hat{\theta}=0}$'' gauge can be found in
 \cite{BdAI,BdAIL03,IB+JI=03,IB+JdA=05}. We notice only few of its properties.

Firstly,  in the WZ gauge (\ref{WZgauge=f}), (\ref{WZgauge=b}) the leading component of supervielbein matrix has a triangular form,
 \begin{eqnarray}\label{WZ0gg2}
E_N{}^A\vert_{\theta =0} = \left(\matrix{ e_\nu^a(x) &
\psi_\nu^{\underline{\alpha}}(x)\cr
0 & \delta_{\breve{\beta}}{}^{\underline{\alpha}} }\right) \qquad \Rightarrow \qquad  E_A{}^N\vert_{\theta =0} = \left(\matrix{e_a^\nu(x) &
- \psi_a^{\breve{\beta}}(x)\cr
0 & \delta_{\underline{\alpha}}{}^{\breve{\beta}}}\right) \; ,
\end{eqnarray}
which implies, in particular, the following relation between the leading component of $T_{ab}{}^{\alpha}$ and the true gravitino field strength ${\cal D}_{[\mu}\psi_{\nu]}^{\alpha}(x)$
\begin{eqnarray}\label{TbbfWZ}
T_{ab}{}^{\alpha}\vert_{\theta =0} & = 2 e_a^\mu e_b^\nu
{\cal D}_{[\mu}\psi_{\nu]}^{\alpha}(x) -  {i\over 4} (\psi_{[a}\sigma_{b]})_{\dot{\beta}}
G^{\alpha\dot{\beta}}\vert_{\theta =0}
 - {i\over 4} (\bar{\psi}_{[a}\tilde{\sigma}_{b]})^{\alpha}
R\vert_{\theta =0} \; .
\end{eqnarray}

Secondly, we would like to comment on symmetries leaving Eqs. (\ref{WZgauge=f})--(\ref{thGauge}) invariant. The WZ gauge (\ref{WZgauge=f}), (\ref{WZgauge=b}), (\ref{WZgauge=w}) is preserved by spacetime diffeomorphisms, local Lorentz symmetry and supersymmetry. Fixing further the gauge (\ref{thGauge}) we break 1/2 of the local supersymmetry {\it on the worldvolume of the supermembrane}. The only restriction on the parameter of the local spacetime supersymmetry $\epsilon^{\alpha}(x)$ is the condition that its pull--back to $W^3$, $\hat{\epsilon}{}^{{\alpha}}:= {\epsilon}{}^{{\alpha}}(\hat{x}(\xi))$, and its complex conjugate $\hat{\bar{\epsilon}}{}^{\dot{\alpha}}:= \bar{\epsilon}^{\dot{\alpha}}(\hat{x}(\xi))
$ are related by
\begin{eqnarray}\label{1/2susy}
\hat{\epsilon}{}^{{\alpha}}= \hat{\bar{\epsilon}}_{\dot{\alpha}}\tilde{\bar{\gamma}}{}^{{\dot{\alpha}\alpha}}\; ,\qquad
\end{eqnarray}
where $\tilde{\bar{\gamma}}{}^{{\dot{\alpha}\alpha}}$ is the supermembrane $\kappa$--symmetry projector (\ref{tg:=}) calculated with $\hat{\theta}(\xi)=0$. Eq. (\ref{1/2susy}) is tantamount to saying that the pull--back of the local supersymmetry parameter to $W^3$ is expressed through the $\kappa$--symmetry parameter of the supermembrane. There are no restrictions on the local supersymmetry parameter outside the supermembrane  worldvolume so that the equations (\ref{1/2susy}) can be understood as the boundary condition imposed on the supersymmetry parameter on the domain wall $W^3$.

\subsection{Current superfields in the WZ$_{\hat{\theta}=0}$ gauge. Current prepotentials and Rarita--Schwinger equation}

In the gauge (\ref{WZgauge=f})--(\ref{1/2susy}),
\begin{eqnarray}\label{hEb=gauge}
\hat{E}{}^a= \hat{e}{}^a= d\hat{x}{}^\mu e_\mu{}^a(\hat{x})    \; ,  \qquad
\hat{E}{}^{\alpha} = \hat{\psi}{}^{\alpha}= d\hat{x}{}^\mu {\psi}_\mu{}{}^{\alpha}(\hat{x})    \; ,  \qquad
\end{eqnarray}
and
\begin{eqnarray}\label{fDdel}
& {\cal D}_\alpha \delta^8(Z-\hat{Z})= {1\over 8} \theta_\alpha\, \bar{\theta}\bar{\theta}\, \delta^4(x-\hat{x}) + \propto \underline{{\theta}}^{\wedge 4}
\, ,  \quad \bar{{\cal D}}_{\dot\alpha} \delta^8(Z-\hat{Z})= - {1\over 8} \bar{\theta}_{\dot\alpha}\, {\theta}{\theta}\, \delta^4(x-\hat{x}) + \underline{{\theta}}^{\wedge 4}
\; ,  \quad \\ \label{fDfDdel}
& {\cal D}^\alpha {\cal D}_\alpha \delta^8(Z-\hat{Z})= - {1\over 4}  \bar{\theta}\bar{\theta}\, \delta^4(x-\hat{x}) + \propto {\theta} \, \bar{\theta}\bar{\theta}
\, ,  \quad \nonumber \\ &  \bar{{\cal D}}_{\dot\alpha} \bar{{\cal D}}{}^{\dot\alpha} \delta^8(Z-\hat{Z})= - {1\over 4} {\theta}{\theta}\, \delta^4(x-\hat{x}) + \propto {\theta}{\theta} \, \bar{\theta}
\; , \quad  \\
\label{fDbfDdel}
& {}[{\cal D}_\alpha ,  \bar{{\cal D}}_{\dot\alpha} ]\delta^8(Z-\hat{Z})= -{1\over 2} \theta_\alpha\, \bar{\theta}_{\dot\alpha}\, \delta^4(x-\hat{x}) +  \propto \underline{{\theta}}^{\wedge 3}
\; ,  \quad
\end{eqnarray}
where $\underline{{\theta}}^{\wedge 4}:= {\theta}{\theta} \, \bar{\theta}\bar{\theta}$
and
$\underline{{\theta}}^{\wedge 3}$ denotes terms proportional to either  ${\theta}{\theta} \, \bar{\theta}$ or ${\theta} \, \bar{\theta} \bar{\theta} $ (or both, which implies $\propto {\theta}{\theta} \, \bar{\theta}\bar{\theta}$). Using these relations and introducing the current pre-potential fields
\begin{eqnarray}\label{Kab(x)=}
{\cal P}_a{}^b(x)&:=& \int\limits_{W^3}  {1\over \hat{e}} *\hat{e}_a \wedge \hat{e}{}^b  \, \delta^4 (x-\hat{x})  \; ,  \qquad
 \\ \label{Kab(x)=}
{\cal P}_a(x) &:=& \int\limits_{W^3}   {1\over \hat{e}} \epsilon_{abcd} \hat{e}{}^b \wedge \hat{e}{}^c \wedge \hat{e}{}^d \,
 \, \delta^4 (x-\hat{x}) = \qquad \nonumber \\ &=& e_a^\mu (x) \int\limits_{W^3}    \epsilon_{\mu\nu\rho\sigma} d\hat{x}{}^\nu \wedge  d\hat{x}{}^\rho \wedge  d\hat{x}{}^\sigma  \,
 \, \delta^4 (x-\hat{x}) \;  , \qquad
\end{eqnarray}
we find that the vector and scalar current superfields (\ref{Ja=NG+WZ}), (\ref{cX=}) have the form
\begin{eqnarray}\label{Ja=gauge}
J_{\alpha \dot{\alpha}} \vert_{\hat{\theta}=0} =  {\theta_\beta\, \bar{\theta}_{\dot{\beta}} \over   8}
(\;  3 {\cal P}_a{}^b(x) {\sigma}^a_{\alpha \dot{\alpha}}\tilde{\sigma}{}_b^{\beta\dot{\beta}}- 2\delta_{\alpha}{}^{\beta} \delta_{\dot{\alpha}}{}^{\dot{\beta}} {\cal P}_b{}^b(x)) - i {({\theta}{\theta}  - \bar{\theta}\bar{\theta})\over 32}   {\sigma}^a_{\alpha \dot{\alpha}} {\cal P}_a(x)
+  \propto \underline{\theta}^{\wedge 3}
\qquad
\end{eqnarray}
and
\begin{eqnarray}\label{cX=gauge}
{\cal X} \vert_{\hat{\theta}=0} &=& - {\theta{\sigma}^a\bar{\theta} \over 16} {\cal P}_a+ i {({\theta}{\theta}  - \bar{\theta}\bar{\theta})\over 16}   {\cal P}_a{}^a(x)
+  \propto \underline{\theta}^{\wedge 3}
\; . \quad
\end{eqnarray}
Using (\ref{Ja=gauge}) and (\ref{cX=gauge}) one can easily check that Eqs. (\ref{DJ=-iDX}) are satisfied at lowest order in $\underline{\theta}$.

One also sees that there is no explicit supermembrane contributions to the Rarita--Schwinger equations of the supergravity--supermembrane interacting system which thus reads
\begin{eqnarray}\label{SG+2p=RS=on}
\epsilon^{\mu\nu\rho\sigma}
e_{\nu}^a (x) {\cal D}_\rho \psi_\sigma^{\alpha}(x)\, \sigma_{a\alpha\dot{\alpha}} =0\; .
\end{eqnarray}

However, such a contribution is actually present in (\ref{SG+2p=RS=on}) implicitly, hidden inside the covariant derivative. Indeed, as indicated by Einstein equation, the bosonic vielbein and the spin connection do contain some contributions from supermembrane.

\subsection{Einstein equation of the supergravity--supermembrane interacting system in the WZ $_{\hat{\theta}=0}$ gauge}

The Einstein equation with supermembrane current contributions can be obtained as leading term in the decomposition  of Eq. (\ref{RRicci}), {\it i.e.}
\begin{eqnarray} \label{RRicci=J}
R_{bc}{}^{ac}\vert_{_{\theta=0}}& =& {1\over 32} ({{\cal D}}^{{\beta}} \bar{{\cal D}}^{(\dot{\alpha}|}
J^{{\alpha}|\dot{\beta})} - \bar{{\cal D}}^{\dot{\beta}} {{\cal D}}^{({\beta}}J^{{\alpha})\dot{\alpha}})\vert_{_{\theta=0}}
\sigma^a_{\alpha\dot{\alpha}}\sigma_{b\beta\dot{\beta}} - {3i\over 64}
(\bar{{\cal D}}\bar{{\cal D}}{\cal X} - {{\cal D}}{{\cal D}}{\cal X})\vert_{_{\theta=0}} \, \delta_b^a + \qquad \nonumber \\ && + {3\over 16} (R\bar{R})\vert_{_{\theta=0}} \delta_b^a\; . \qquad
\end{eqnarray}
The first two terms in the {\it r.h.s.} of Eq. (\ref{RRicci=J}) can be easily calculated from Eqs. (\ref{Ja=gauge}), (\ref{cX=gauge}), while the last term, in the light of that the scalar superfield equation has the form of Eq. (\ref{R-bR=cX}), is expressed in terms of $(R+\bar{R})^2\vert_{_{\theta=0}}$  and requires a separate study. As an intermediate resume let us fix that
\begin{eqnarray} \label{RRicci=Jg1}
R_{bc}{}^{ac}\vert_{_{\theta=0\; , \; \hat{\theta}=0}}& =& - {3\over 32} \, T_2\, \left( {\cal P}_b{}^a(x) - {1\over 2} \delta_b^a  {\cal P}_c{}^c(x)\right) + {3\over 64} (R+\bar{R})^2\vert_{_{\theta=0}} \delta_b^a\; . \qquad
\end{eqnarray}
The last term is the square of $(R+\bar{R})\vert_{_{\theta=0}}$ which, as a result of (\ref{R-bR=cX}), obeys the equation
\begin{eqnarray} \label{d(R+bR)=}
\partial_\mu (R+\bar{R})\vert_{_{\theta=0}}= {T_2\over 16} \int\limits_{W^3}    \epsilon_{\mu\nu\rho\sigma} d\hat{x}{}^\nu \wedge  d\hat{x}{}^\rho \wedge  d\hat{x}{}^\sigma  \,
 \, \delta^4 (x-\hat{x}) \;  . \qquad
\qquad
\end{eqnarray}
The solution of this equation can be written in the form
\begin{eqnarray} \label{(R+bR)=c+int}
R(x)+\bar{R}(x)= 8c +{ T_2\over 16} \int\limits^x_{x_0} d\tilde{x}^\mu \int\limits_{W^3}     \epsilon_{\mu\nu\rho\sigma} d\hat{x}{}^\nu \wedge  d\hat{x}{}^\rho \wedge  d\hat{x}{}^\sigma  \,
 \, \delta^4 (\tilde{x}-\hat{x})  \; ,
\end{eqnarray}
where $c$ is an arbitrary   constant which corresponds to the value of $(R+\bar{R})$ at the spacetime point $x_0^\mu$ providing the lower limit of the integral in the second term, $c=(R(x_0)+\bar{R}(x_0))/8$.

One can easily check that
\begin{eqnarray}\label{Theta=}
\Theta (x,x_0|\hat{x}):=
\int\limits^x_{x_0} d\tilde{x}^\mu \int\limits_{W^3}     \epsilon_{\mu\nu\rho\sigma} d\hat{x}{}^\nu \wedge  d\hat{x}{}^\rho \wedge  d\hat{x}{}^\sigma  \,
 \, \delta^4 (\tilde{x}-\hat{x}) \; ,  \qquad
\end{eqnarray}
entering the second term in the {\it r.h.s.} of (\ref{(R+bR)=c+int}), obeys
\begin{eqnarray}\label{dTheta=1}
\partial_\mu \Theta (x,x_0|\hat{x})= \int\limits_{W^3}     \epsilon_{\mu\nu\rho\sigma} d\hat{x}{}^\nu \wedge  d\hat{x}{}^\rho \wedge  d\hat{x}{}^\sigma  \,\, \delta^4 (x-\hat{x})\; . \qquad
\end{eqnarray} Furthermore, using a convenient local frame in the neighborhood of the worldvolume, one can check that $\Theta (x,x_0|\hat{x})$ vanishes if the points $x^\mu$ and ${x}^\mu_0$ are on the same side of spacetime with respect to the domain wall provided by the supermembrane worldvolume while it is equal to $\pm 1$  if these points belongs to the different branches  of the spacetime separated by this domain wall. This is  to say that Eq. (\ref{Theta=}) defines a counterpart of the Heaviside  step function associated to the direction orthogonal to the supermembrane worldvolume. The last statement about association implies that
$\Theta (x,x_0|\hat{x})$ is a functional of the supermembrane coordinate function $\hat{x}^\mu(\xi)$.
Furthermore, as in the case of the standard Heaviside  step function $\Theta (y)$, we can use
 \footnote{In the case of standard standard Heaviside  step function this is equivalent to setting the indefinite value $\Theta (0)$ equal to $1/2$. Indeed,
calculating the derivative $\partial_y (\Theta(y)\Theta (y))= 2 \Theta(y)\delta (y)= 2 \Theta(0)\delta (y)$  we find that this coincides with $\partial_y \Theta(y)= \delta (y)$ when  $\Theta (0)=1/2$.  In our case $\Theta (x,x_0|\hat{x})$ is the counterpart of either $+\Theta(y)$ or $-\Theta(y)$ so that $(\Theta (x,x_0|\hat{x}))^2= \pm \Theta (x,x_0|\hat{x})$. However, by a suitable choice of the location of the point $x_0$ with respect to $W^3$ one can always arrive at the situation with nonnegative $\Theta (x,x_0|\hat{x})$. Below for simplicity we assume this choice is made. }
 $(\Theta (x,x_0|\hat{x}))^2= \Theta (x,x_0|\hat{x})$.

Thus our solution of the auxiliary field equations (\ref{(R+bR)=c+int}) can be written as ({\it cf. } \cite{Aurilia:1978qs})
\begin{eqnarray} \label{(R+bR)=c+T2}
R(x)+\bar{R}(x)= 8c + { T_2\over 16} \Theta (x,x_0|\hat{x})\;
\end{eqnarray}
and implies that Eq. (\ref{RRicci=Jg1}) reads
\begin{eqnarray} \label{RRicci=K+Hev}
R_{bc}{}^{ac}(x)& =& - {3T_2\over 32} \left( {\cal P}_b{}^a(x) - {1\over 2} \delta_b^a  {\cal P}_c{}^c(x)\right)  + 3
\delta_b^a \, \left( c^2 + \left(\left({T_2\over 128} +  c \right)^2 -  c^2 \right) \Theta (x,x_0|\hat{x})\right)\; , \qquad
\end{eqnarray}
where ${\cal P}_b{}^a(x)$ is the singular contribution  defined in (\ref{Kab(x)=}).

\subsection{Cosmological constant generation in the interacting system and its ``renormalization'' due to supermembrane}

Let us analyze the supermembrane contribution to the Einstein equations.
These can be separated in two classes, one containing singular contributions and the other containing regular contributions proportional to $T_2$.

Being a bit more provocative one can say about three classes, counting also  the contribution proportional to square of the arbitrary integration constant $c$, as far as this comes from the auxiliary field sector of the special minimal supergravity, the off--shell formulation which is 'elected' by the supermembrane. As we have already commented in sec. 2.4. and in \cite{IB+CM=2011}, the supermembrane can exist in a background of a generic minimal  supergravity, however the supermembrane interaction with dynamical supergravity  requires this to be special minimal supergravity. This in its turn, even in the absence of any matter (neither of the field theoretical type nor of branes), produces Einstein equations with a cosmological constant generated dynamically. Then this cosmological constant proportional to the square of the above  arbitrary integration constant $c$ should also be considered as a(n indirect) contribution of the supermembrane to the Einstein equation.

To be more concrete, Eq. (\ref{RRicci=K+Hev}) can be written in the form $R_{bc}{}^{ac}(x) =  3 c^2
\delta_b^a + \propto T_2$,
\begin{eqnarray} \label{RRicci=ta2+T2t}
R_{acb}{}^{c}(x)& =&
\eta_{ab} \, 3c^2 + T_2  \left( {\cal T}_{ab}^{sing}(x) + {\cal T}_{ab}^{reg}(x)\right)\; . \qquad
\end{eqnarray}
 When $T_2$ is set to zero, it contains a nonvanishing cosmological constant contribution with $\Lambda =- {3c^2}$. This (AdS-type) cosmological constant is generated dynamically as far as it is proportional to the (minus) square of the arbitrary integration constant $c$ which is inevitable in the special minimal supergravity equations due to its auxiliary field structure (see \cite{Ovrut:1997ur} and \cite{IB+CM=2011} for references and more discussion). In its turn, special minimal supergravity, and not generic minimal supergravity can be included into the action of the supergravity--supermembrane interacting system. In this sense the cosmological constant generated dynamically is the first 'relict' contribution from the supermembrane to the Einstein equation of the interacting system.

The second type of the supermembrane contributions to the {\it r.h.s.} of the Einstein equation are singular terms $\propto {\cal P}_c{}^d(x)$ (\ref{Kab(x)=}),
\begin{eqnarray} \label{T2sing=}
 {\cal T}_{ab}^{sing}(x)&=& - T_2 {3\over 32} \left({\cal P}_{ba}(x) - {1\over 2} \eta_{ba}  {\cal P}_c{}^c(x)\right)= \qquad \nonumber \\
 &=&- {3T_2\over 32} \int\limits_{W^3}  {1\over \hat{e}} *\hat{e}_a \wedge \hat{e}{}_b  \, \delta^4 (x-\hat{x}) + {3T_2\over 64} \eta_{ba}  \int\limits_{W^3}  {1\over \hat{e}} *\hat{e}_c \wedge \hat{e}{}^c  \, \delta^4 (x-\hat{x})
 \;  \qquad
\end{eqnarray}
which are expected when (super)gravity interact with supermembrane.

In the  third type we collect the regular  supermembrane contributions which are proportional to the supermembrane tension,
\begin{eqnarray} \label{T2reg=}
 {\cal T}_{ab}^{reg}(x)&=& \eta_{ab} {\cal T}^{reg}(x) \; , \qquad  {\cal T}^{reg}(x) =
+ {3T_2\over 64}
\left({T_2\over 256} + {c} \right)   \Theta (x,x_0|\hat{x})\; . \qquad \;  \qquad
\end{eqnarray}

To appreciate the role of this contribution it is instructive to consider the Einstein equation in two pieces of the spacetime separated by the supermembrane worldvolume. Let us denote the half-space where $\Theta (x,x_0|\hat{x})=1$ by $M^4_+$ and the half-space where $\Theta (x,x_0|\hat{x})=0$ by $M^4_-$. Then the singular terms  (\ref{T2sing=}) do not contribute and the Einstein equation reads
\begin{eqnarray} \label{RRicci+=}
M^4_+\; : \qquad R_{acb}{}^{c}(x)& =&   3
\eta_{ab} \, \left({T_2\over 128} +  c \right)^2\; . \qquad
\\ \label{RRicci-=}
M^4_-\; : \qquad R_{acb}{}^{c}(x)& =&   3
\eta_{ab} \, c^2 \; . \qquad
\end{eqnarray}

An evident observation is that, in the general case,
the  cosmological constants in  different branches of spacetime separated
by the worldvolume $W^3$ are different.

One also notices that the cosmological constants in $M^4_+$ and $M^4_-$ coincide if
$c=-{T_2\over 256}$. However, as far as $c$ is an arbitrary integration constant, fixing its value is equivalent to imposing a kind of boundary conditions and we do not see any special reason to chose such boundary conditions in such a way that $c=-{T_2\over 256}$. Rather we should allow a generic value of  $c$ and thus accept that the cosmological constant takes different values in the branches of spacetime separated by the supermembrane worldvolume.

Notice that the solution of the Einstein equation describing membranes separating two $AdS_5$ spaces with different values of cosmological constants were studied in \cite{Gogberashvili:1998iu}, as a Brane world alternative to the dark matter, and  \cite{JMMS+=2001} in relation with the hypothesis on possible change of signature in the Brane World models. See also \cite{shellUni} for the related studies. In the bosonic perspective the appearance of different cosmological constants on the different sides of a domain wall interacting with gravity and a 3--form gauge field was known from \cite{Aurilia:1978qs}, where it was used as a basis for a bag model for hadrons, and from  \cite{Brown:1988kg} where this effect was proposed as a mechanism for damping the cosmological constant. Our present study indicates that the result on the different values of cosmological constant on the different sides of the supermembrane domain wall is an imminent consequence of the dynamics of the supersymmetric interacting system of the supermembrane and dynamical $D=4$  ${\cal N}=1$ supergravity.

\section{On supersymmetric solutions of the interacting system equations}
\label{solutions}
When searching for purely bosonic supersymmetric solutions, setting
$\psi_\mu^\alpha=0$, one studies the Killing spinor equations, which appears as the conditions of supersymmetry preservation, $\delta_\epsilon \psi_\mu^\alpha=0$. When starting from superfield formulation of supergravity, $\delta_\epsilon \psi_\mu^\alpha$ can be calculated with the use of superspace Lie derivative, this is to say $\delta_\epsilon \psi_\mu^\alpha= D_\mu\epsilon^{\alpha} + (E_\mu^C \epsilon^{\underline{\beta}}T_{\underline{\beta} C}{}^\alpha )\vert_{\theta =0} $.
Hence, in a generic off-shell $D=4$ ${\cal N}=1$  minimal supergravity background the Killing equations are
\begin{eqnarray}\label{Killing:=De+}
 {D}\epsilon^{\alpha} + {i\over 8} e^c (\epsilon
\sigma_c\tilde{\sigma}_d)_{\beta} {}^{\alpha}\; G^d\vert_{\theta =0}   +{i\over 8} e^c
\; (\bar{\epsilon}\tilde{\sigma}_{c}){}^{\alpha} \;R\vert_{\theta =0}=0  \; \qquad
\end{eqnarray}
and the complex conjugate equation. Using the superfield equations of motion (\ref{Ga=Ja}), (\ref{R-bR=cX}), the explicit form of the current superfields in the WZ$_{\hat{\theta}=0}$ gauge, Eqs. (\ref{Ja=gauge}), (\ref{cX=gauge}), and Eq. (\ref{(R+bR)=c+T2}), we find that the Killing equation (\ref{Killing:=De+}) reads
\begin{eqnarray}\label{De=ta+}
{D}\epsilon^{\alpha} + {i\over 2} e^a
\; (\bar{\epsilon}\tilde{\sigma}_{a}){}^{\alpha}  \left( c + { T_2\over 128} \Theta (x,x_0|\hat{x})\right)=0\; .\qquad
\end{eqnarray}
We can split this on two Killing equations valid in two different branches of spacetime separated by the  supermembrane worldvolume,
\begin{eqnarray}\label{M-=Killing}
M_-^4\, &:& \qquad {D}\epsilon^{\alpha} + {i\over 2} e^a
\; (\bar{\epsilon}\tilde{\sigma}_{a}){}^{\alpha} \, c =0\; , \qquad \\ \label{M+=Killing}
M_+^4\, &:& \qquad {D}\epsilon^{\alpha} + {i\over 2} e^a
\; (\bar{\epsilon}\tilde{\sigma}_{a}){}^{\alpha}  \left( c + { T_2\over 128} \right)=0\; .\qquad
\end{eqnarray}
The supersymmetry parameter should also obey the boundary conditions (\ref{1/2susy}) on the worldvolume $W^3$, which is  the common boundary of $M_+^4$ and $M_-^4$,
\begin{eqnarray}\label{W3=1/2susy}
W^3= \pm \partial M_\pm^4\; : \qquad  \hat{\epsilon}{}^{{\alpha}}= \hat{\bar{\epsilon}}_{\dot{\alpha}}\tilde{\bar{\gamma}}{}^{{\dot{\alpha}\alpha}}\; ,\qquad
\hat{\epsilon}{}^{{\alpha}}:= {\epsilon}{}^{{\alpha}}(\hat{x}(\xi)) \; , \quad \hat{\bar{\epsilon}}{}_{\dot{\alpha}}:= \bar{\epsilon}_{\dot{\alpha}}(\hat{x}(\xi)) \; .  \qquad
\end{eqnarray}

The detailed study of these system of Killing spinor equations and the search for the supersymmetric solutions  of the interacting system equations on their basis is an interesting subject for future.
An intriguing question is whether the supersymmetric solutions of the equations of the interacting system exist in the generic case of arbitrary $c$ corresponding to different values of cosmological constants on different sides of the supermembrane worldvolume, or supersymmetry selects some particular values of the constant $c$.
Presently we can state that if  obstructions existed, they would occur due to the singular terms with support on the worldvolume $W^3$, while the mere fact of different values of cosmological constant on the branches of spacetime situated on the different sides of $W^3$ does not prohibit supersymmetry. Indeed, let us study the integrability conditions for the Killing spinor equations in $M_\pm^4$. Applying the exterior covariant derivatives to  Eqs. (\ref{M-=Killing}) and (\ref{M+=Killing}) and using the Ricci identities $DD\epsilon^{\alpha}=- {1\over 4} R^{ab} \epsilon^{\beta} {\sigma}_{ab\,\beta}{}^{\alpha}$ and the equations complex conjugate to (\ref{M-=Killing}) and (\ref{M+=Killing}), we find
\begin{eqnarray}\label{M-=DKilling}
M_-^4\, &:& \qquad  R^{ab} \epsilon^{\beta} {\sigma}_{ab\,\beta}{}^{\alpha} =  {1\over 4}|c|^2 e^d\wedge e^c \,  \epsilon^{\beta}{\sigma}_{cd\,\beta}{}^{\alpha} \; ,\qquad \\ \label{M+=DKilling}
M_+^4\, &:& \qquad  R^{ab} \epsilon^{\beta} {\sigma}_{ab\,\beta}{}^{\alpha} =  {1\over 4}\left| c + { T_2\over 128} \right|^2 e^d\wedge e^c \,  \epsilon^{\beta}{\sigma}_{cd\,\beta}{}^{\alpha}
\; . \qquad
\end{eqnarray}
If we search for a purely bosonic  solution preserving all the supersymmetry in $M_-^4$ and $M_+^4$,  Eqs. (\ref{M-=DKilling}) and (\ref{M+=DKilling}) should be obeyed for an arbitrary $\epsilon^\alpha$. This implies
\begin{eqnarray}\label{RM-=susy}
M_-^4\, &:& \qquad R_{cd}{}^{ab}=    {1\over 2}|c|^2 {\delta}_{[c}{}^a{\delta}_{d]}{}^b \; ,\qquad \\ \label{RM+=susy}
M_+^4\, &:& \qquad R_{cd}{}^{ab}={1\over 2} \left| c + { T_2\over 128} \right|^2 {\delta}_{[c}{}^a{\delta}_{d]}{}^b
\; , \qquad
\end{eqnarray}
{\it i.e.} that $M_\pm^4$ are AdS spaces with apparently different cosmological constants. One can easily check that (\ref{RM-=susy}) and (\ref{RM+=susy}) solve our equations of motion  (\ref{RRicci-=}) and (\ref{RRicci+=}) and thus describe the completely supersymmetric solution of the system of the supergravity equations of the interacting system (at least) when these are considered modulo singular terms with the support on $W^3$.

Let us stress that such a system of equations does contain the supermembrane contributions: not only an indirect, which comes from an arbitrary cosmological constant generated dynamically due to the structure of the supergravity auxiliary fields imposed by the supergravity interaction with supermembrane (see \cite{Ovrut:1997ur} and also \cite{Aurilia:1978qs,OS80,Aurilia:1980xj,Duff:1980qv}), but also {\it direct}, which is a shift of cosmological constant on one of the sides of the brane worldvolume on the value which is proportional to the supermembrane tension (see \cite{Aurilia:1978qs,Brown:1988kg}).
Furthermore, although preserving all 4 supersymmetries in  $M_-^4$ and $M_+^4$, when considered as a solution of the equations of interacting system, Eqs. (\ref{RM-=susy}) and (\ref{RM+=susy}) describe the 1/2 BPS state, {\it i.e.} the state preserving 1/2 of the supersymmetry. Indeed, when considering the interacting system we have to restrict the local supersymmetry parameter by the boundary conditions (\ref{W3=1/2susy}) on $W^3$ and these clearly break 1/2 of the supersymmetry on $W^3$.

\section{Conclusion}

In this paper  we have derived the complete set of spacetime component  equations of motion for the interacting system of dynamical $D=4$ ${\cal N}=1$  supergravity and supermembrane.
To this end we obtain the superfield equations of motion by varying the action given by the sum of the Wess--Zumino superfield supergravity action \cite{WZ77,WZ78} and the action for supermembrane \cite{AGIT=88} in curved $D=4$ ${\cal N}=1$ superspace. To obtain the spacetime component equations of motion we have used the Wess--Zumino gauge supplemented by partial gauge fixing of the local  supersymmetry on the  supermembrane worldvolume $W^3$ which is achieved by setting to zero the Goldstone fermion of the supermembrane, $\hat{\theta}{}^{\underline{\alpha}}(\xi)=0$ (see \cite{BdAIL03,IB+JI=03}). We have shown that the  supermembrane current superfields simplify drastically in this  ''WZ$_{\hat{\theta}=0}$ gauge''.

Our interacting system includes  the Grisaru--Siegel--Gates--Ovrut--Waldram   special minimal supergravity \cite{Ovrut:1997ur}, the Wess--Zumino type superspace formulation of which was developed in \cite{IB+CM=2011}. The Einstein equation of this supergravity includes a cosmological constant generated dynamically, {\it i.e.} expressed through (minus square of) an arbitrary integration constant $c$ \cite{Ovrut:1997ur,IB+CM=2011} (see \cite{Aurilia:1978qs,Aurilia:1980xj,Duff:1980qv} for earlier study of this effect).

When the special minimal supergravity interacts with supermembrane, this latter produces a kind of renormalization of this cosmological constant, making its value different in the branches of spacetime separated by the supermembrane worldvolume. Namely, the Einstein equation acquires as well the supermembrane energy momentum tensor contributions.
Besides singular contributions with the support on $W^3$, the supermembrane energy---momentum tensor contains also some nonsingular contributions proportional to a covariant version of the Heaviside step function $\Theta (x,x_0|\hat{x})$, which is equal either to unity, or to zero. To be more precise, the domain wall of the supermembrane worldvolume $W^3$, which can be defined by the equation $x^\mu=\hat{x}^\mu(\xi)$ separates the spacetime into two branches, $M^4_+$ and $M^4_-$, and the above  generalized Heaviside function  $\Theta (x,x_0|\hat{x})$, is equal to unity in (say) $M^4_+$  and is equal to   zero in $M^4_-$. Then, the regular supermembrane contribution proportional to  $\Theta (x,x_0|\hat{x})$ makes the cosmological constant in $M^4_+$ and $M^4_-$ different.

In the purely bosonic interacting system of gravity, membrane and antisymmetric 3-form gauge field, such effect was described in  \cite{Aurilia:1978qs,Brown:1988kg}. The solution of the Einstein equations describing a domain wall separating two branches of AdS space with different cosmological constants have been considered in literature \cite{Gogberashvili:1998iu,JMMS+=2001,shellUni}, in particular in the context of the brane world alternative to the dark matter \cite{Gogberashvili:1998iu}. As we have shown in Sec. \ref{solutions}, such a configuration provides  a supersymmetric solution of the system of our supergravity equations considered outside the supermembrane  worldvolume $W^3$.

Let us stress that generically the difference of the values of cosmological constants in $M^4_+$ and $M^4_-$ is proportional to the supermembrane tension $T_2$, while its basic  value is determined by an arbitrary constant $c$ and is independent on $T_2$. Only for the special case $c=-{T_2\over 256}$ for which  the cosmological constant is equal in the branches of spacetime separated by $W^3$ this value of cosmological constant is proportional to $T_2^2$.  However, presently we do not see any reason to prefer this solution of the interacting system equations. The supersymmetric solution for the case $c=-{T_2\over 256}$, when the values of the cosmological constant in $M^4_+$ and $M^4_-$ coincide  and are proportional to the supermembrane tension, can be found in \cite{Ovrut:1997ur}.

The consistent accounting for the singular terms with the support on $W^3$ and the search for the solutions of interacting system equations preserving some part ($\leq 1/2$)  of supersymmetry is an interesting subject for future study.

Another natural way to develop the results of present paper is to study the superfield Lagrangian description of the more general $D=4$ interacting system including  supermembrane, supergravity and matter multiplets and to compere its result with the spacetime component study of \cite{Tomas+}.

\bigskip

{\it Acknowledgments}. The authors are thankful Jos\'e M.M. Senovilla and Paul K. Townsend for useful discussions and communications.  The partial support by the research grants FIS2008-1980 from the Spanish MICINN (presently MIEyC)  and by the Basque Government Research Group Grant ITT559-10 is greatly acknowledged.

\section*{Appendix: On admissible  variations of superfield supergravity}

The admissible variations of supervielbein of minimal supergravity superspace  read \cite{WZ78,BdAIL03}
\begin{eqnarray}\label{varEa} \delta E^{a} & =& E^a (\Lambda (\delta ) + \bar{\Lambda} (\delta ))
 - {1\over 4} E^b \tilde{\sigma}_b^{ \dot{\alpha} {\alpha} }
[{\cal D}_{{\alpha}}, \bar{{\cal D}}_{\dot{\alpha}}] \delta H^a +
 i E^{\alpha} {\cal D}_{{\alpha}}\delta H^a   - i \bar{E}^{\dot{\alpha}}\bar{{\cal
D}}_{\dot{\alpha}} \delta H^a \; , \\ \label{varEal}
 \delta E^{\alpha} & = & E^a \Xi_a^{\alpha}(\delta ) +
E^{\alpha} \Lambda (\delta ) + {1\over 8} \bar{E}^{\dot{\alpha}} R \sigma_a{}_{\dot{\alpha}}{}^{\alpha}
\delta H^a \; ,  \end{eqnarray}
where
\begin{eqnarray}
\label{2Lb+*Lb} & 2\Lambda (\delta ) + \bar{\Lambda} (\delta )  = {1\over 4} \tilde{\sigma}_a^{
\dot{\alpha} {\alpha} } {\cal D}_{{\alpha}} \bar{{\cal D}}_{\dot{\alpha}}\delta H^a + {1\over 8} G_a
\delta H^a +   3 ( {\cal D}{\cal D}- \bar{R})\delta {\cal U} \;
 \end{eqnarray}
and the explicit expression for $\Xi_a^{\alpha}(\delta )$ in (\ref{varEal}) can be found in \cite{BdAIL03}.

The variation of the closed 4--form (\ref{H4=HL+HR}), (\ref{H4L}) reads \cite{IB+CM=2011}
\begin{eqnarray} \label{vH4=}  \delta H_4 &=& {1\over 2} E^b\wedge  E^\alpha \wedge  E^\beta\wedge  E^\gamma
\sigma_{ab\; (\alpha\beta} D_{\gamma )}\delta H^a  - {1\over 2} E^b\wedge  E^\alpha \wedge  E^\beta\wedge   \bar{E}{}^{\dot{\gamma}}
\sigma_{ab\; \alpha\beta} \bar{D}_{\dot\gamma}\delta H^a + c.c. - \qquad
 \nonumber \\ && - {i\over 2} E^b\wedge E^a \wedge E^\alpha \wedge E^\beta
\left(\sigma_{ab\; \alpha\beta}\left(2\Lambda(\delta) + \bar{\Lambda}(\delta)\right)+ {1\over 4}\sigma_{c[a|\; \alpha\beta}\tilde{\sigma}_{|b]}{}^{\dot{\gamma}\gamma}[D_\gamma, \bar{D}_{\dot{\gamma}}]\delta H^c
\right) + c.c. + \nonumber \\ && + {i\over 16} E^b\wedge  E^a \wedge  {E}{}^{\alpha} \wedge \bar{E}{}^{\dot\beta} (R\sigma_{ab}\tilde{\sigma}_{c} -\bar{R}\sigma_c\tilde{\sigma}_{ab})_{\alpha\dot\beta}\delta H^c + \propto  \; E^c \wedge E^b\wedge E^a  \; .
\end{eqnarray}
The conditions of that
$\delta H_4$ can be expressed in terms of the variation of the 3--form potential $\delta C_3$,
 \begin{eqnarray} \label{vH4=dvC3}
\delta H_4 = d(\delta C_3)
 \;  \quad
\end{eqnarray}
with $\delta C_3$ decomposed on the basic covariant 3--forms, as in Eq. (\ref{vC3:=}), restrict the set of independent
variations by  \cite{IB+CM=2011}
\begin{eqnarray}
 \label{DDcU=vV+}
({\cal D}{\cal D}- \bar{R})\delta {\cal U} = {1\over 12}( {\cal D}{\cal D}- \bar{R})\left(i\delta {V} + {1\over 2}  \bar{{\cal D}}_{\dot{\alpha}}  \delta \bar{{\kappa}}{}^{\dot{\alpha}} \right)
\; .     \;
\end{eqnarray}
This is equivalent to
\begin{eqnarray}
 \label{vcU=vV+K}
\delta {\cal U} = {i\over 12}\delta {V} + {1\over 24}  \bar{{\cal D}}_{\dot{\alpha}}  \delta \bar{{\kappa}}{}^{\dot{\alpha}}+ {i\over 24} {\cal D}_{{\alpha}} \delta {{\nu}}{}^{{\alpha}}\; ,   \qquad
\end{eqnarray}
where  $\delta {{\nu}}{}^{{\alpha}}$ is an additional independent  variation; this however does not contribute to $({\cal D}{\cal D}-\bar{R})\delta {\cal U}$ and, hence, to the variations of supergravity potentials.

Factoring out the gauge transformations, we can write the variation $\delta C_3$, which produces (\ref{vH4=}) through (\ref{vH4=dvC3}),  in the form (\ref{vC3:=}) with \cite{IB+CM=2011}
\begin{eqnarray} \label{bff*f=}
\beta_{\alpha\beta\gamma}(\delta)=0= \beta_{\alpha\beta\dot{\gamma}}(\delta)\; , \quad \beta_{{\alpha}\dot{\beta} a}(\delta)  = i\sigma_{a\alpha\dot{\beta}}\delta V \;
\end{eqnarray}
and
\begin{eqnarray} \label{bffb=}
\beta_{\alpha\beta a}(\delta) &=&  - \sigma_{ab\; \alpha\beta} (\delta H^b + \tilde{\sigma}^{b\gamma \dot{\gamma}}D_{\gamma }\delta\bar{\kappa}_{\dot{\gamma}})\; ,    \qquad
 \\ \label{bfbb=}
\beta_{\alpha ab}(\delta) &=& {1\over 2}\epsilon_{abcd}\sigma^c_{\alpha\dot\alpha}\bar{D}{}^{\dot\alpha }\delta H^d +{1\over 2} \sigma_{ab\; \alpha}{}^{\beta} D_\beta \delta V  -
{i\over 4} \tilde{\sigma}_{ab}{}^{\dot{\beta}}{}_{\dot{\gamma}} \bar{D}_{\dot\beta}D_\alpha \bar{\kappa}{}^{\dot{\gamma}}  + {i\over 4} \sigma_{ab\; \alpha}{}^{\beta} \bar{D}_{\dot\beta}D_\beta \bar{\kappa}{}^{\dot{\beta}}   \; ,    \qquad \\
\label{bbbb=}
\beta_{abc}(\delta)  &=& {i\over 8} \epsilon_{abcd}\left( \left({\bar{\cal D}}{\bar{\cal D}}-
{1\over 2}R \right)\delta H^d - c.c. \right) + \nonumber \quad \\   &+& {1\over 4} \epsilon_{abcd}G^d\delta V + {1\over 8} \epsilon_{abcd}\tilde{\sigma}{}^{d\dot{\gamma}\gamma}
[{\cal D}_{\gamma}, {\bar{\cal D}}_{\dot{\gamma}}]\delta V
 - {i\over 16}\epsilon_{abcd}\tilde{\sigma}{}^{d\dot{\gamma}\gamma}\left(\left({\cal D}{\cal D}+{5\over 2}\bar{R}\right) {\bar{\cal D}}_{\dot{\gamma}} {\kappa}_{{\gamma}}- c.c.
  \right)\, .  \qquad
\end{eqnarray}

The variation of the {\it special} minimal supergravity action reads
\begin{eqnarray}\label{vSGsf=sK}  \delta S_{SG} &=&  {1\over 6} \int
d^8Z E\;  \left[ G_a \; \delta H^a + (R-\bar{R}) i\delta {V}  \right] - \nonumber \\ &&
 - {1\over 12} \int
d^8Z E\; \left(R {\cal D}_{{\alpha}} \delta {{\kappa}}{}^{{\alpha}}  +
\bar{R} \bar{{\cal D}}_{\dot{\alpha}}  \delta \bar{{\kappa}}{}^{\dot{\alpha}}\right)
 \; . \qquad
\end{eqnarray}
Notice that the variations $\delta {{\kappa}}{}^{{\alpha}}$ and its complex conjugate (c.c.) result in equations ${\cal D}_{{\alpha}}R=0$ and its  c.c., which are satisfied identically due the minimal (Eq. (\ref{SGeqmR})) or special minimal supergravity equations of motion (Eq. (\ref{SGeqmR+*=0})). In the WZ$_{\hat{\theta}=0}$ gauge (\ref{WZgauge=f})--(\ref{thGauge}) it is also relatively easy to check that $\delta {{\kappa}}{}^{{\alpha}}$ does not produce any independent equation for the physical fields of the interacting system. This observation allowed us to simplify the discussion in the main text by neglecting the existence $\delta {{\kappa}}{}^{{\alpha}}$ variation.

 \end{document}